\documentclass[12pt]{article}

\usepackage[body={17.5cm, 23.1cm},right=2cm]{geometry}

\usepackage{mathtools}

\usepackage[cbgreek]{textgreek}

\usepackage{slashed}

\usepackage{color}
\usepackage{graphicx}
\usepackage{epsf}
\usepackage{graphicx,epsfig}
\usepackage[mathscr]{euscript}
\usepackage{pbox}
\usepackage{float}
\usepackage{amsmath}
\usepackage{amssymb}
\usepackage{epsfig}
\usepackage{cite}
\usepackage{color,colordvi}
\newcommand{\be}{\begin{eqnarray}}
\newcommand{\ee}{\end{eqnarray}}
\newcommand{\bi}{\begin{itemize}}
\newcommand{\ei}{\end{itemize}}

\newcounter{hran}

\def\MSbar{\relax\ifmmode\overline{\rm MS}\else{$\overline{\rm MS}${ }}\fi}
\def\del{\partial}
\def\ta{\text{\textalpha}}

\begin{document}\thispagestyle{empty}

\vspace{0.5cm}

\def\thefootnote{\arabic{footnote}}
\setcounter{footnote}{0}

\def\s{\sigma}
\def\nn{\nonumber}
\def\p{\partial}
\def\ls{\left[}
\def\rs{\right]}
\def\lc{\left\{}
\def\rc{\right\}}
\def\S{\Sigma}
\def\l{\lambda}

\newcommand{\N}{{\cal{N}}}
\newcommand{\beq}{\begin{equation}}
\newcommand{\eeq}[1]{\label{#1}\end{equation}}
\newcommand{\bea}{\begin{eqnarray}}
\newcommand{\eea}[1]{\label{#1}\end{eqnarray}}

\renewcommand{\be}{\begin{eqnarray}}
\renewcommand{\ee}{\end{eqnarray}}
\renewcommand{\th}{\theta}
\newcommand{\bth}{\overline{\theta}}

\hspace*{8cm}
CERN-TH-2020-005, MPP-2020-1, 
LMU-ASC 01/20

\vspace*{1.2cm}

\begin{center}

{\Large \bf 
Aspects of Conformal Supergravity
}
\\[1.5cm]
{\large \bf Sergio Ferrara$^{a,b}$,   \, Alex Kehagias$^{d}$,\, 
Dieter L\"ust$^{e,f}$}

\vspace{0.5cm}

{\it

\vskip-0.3cm
\centerline{ $^{\textrm{a}}$ CERN, Theory Department,}
\centerline{1211 Geneva 23, Switzerland}
\medskip
\centerline{ $^{\textrm{b}}$ INFN, Laboratori Nazionali di Frascati,}
\centerline{Via Enrico Fermi 40, 00044 Frascati, Italy}
\medskip
\centerline{ $^{\textrm{c}}$ Physics Division, National Technical University of Athens}
\centerline{ 15780 Zografou Campus, Athens, Greece}
\medskip
\centerline{ $^{\textrm{d}}$ Arnold--Sommerfeld--Center for Theoretical Physics,}
\centerline{Ludwig--Maximilians--Universit\"at, 80333 M\"unchen, Germany}
\medskip
\centerline{$^{\textrm{e}}$ Max--Planck--Institut f\"ur Physik,
Werner--Heisenberg--Institut,}
\centerline{ 80805 M\"unchen, Germany}

}
\end{center}
\vspace{.8cm}
\vspace{.9cm}

\begin{center}
{\small  \noindent \textbf{Abstract}} \\[0.5cm]
\end{center}
\noindent 
{\small
 In these lectures we discuss ${\cal N}$-extended conformal supergravity and its spectrum in four dimensions. These theories can be considered as the massless limit of Einstein-Weyl 
 supergravity and by taking into account their enhanced gauge symmetries, we derive  their massless spectrum, which in general contains a dipole-ghost graviton multiplet and  an ${\cal N}$-fold tripole-ghost gravitino multiplet.  }
\vskip 2.5cm

\def\thefootnote{\arabic{footnote}}
\setcounter{footnote}{0}
\vfill
\vskip.2in
%
%
%
%
\line(1,0){250}\\
{\footnotesize {To appear in the Proceedings of the 2019 Erice Intl School of Subnuclear Physics, 57th Course.
}}
%
%




\baselineskip= 19pt
\newpage

\section{Introduction}

We would like to report in these lectures on some aspects of the truly ``missing (super)symmetry component of Space and Time" (as phrased by G. 't Hooft \cite{th}) which is the dynamical theory of {\it conformal supergravity}
(or superconformal gravity). The latter is the (local) supersymmetric extension of Weyl gravity and it  should be regarded as the gauge theory of the superconformal algebra.  It   was  pioneered in \cite{FK1,K2} and its quantum structure was explored initially  in \cite{FTs1}. The  fourth order equations of motion of Weyl gravity make the appearence of ghost-like states (of helicity $\pm2,\pm1$) \cite{Stelle} together with ordinary massless graviton, all together making a (massless) dipole (helicity $\pm2$) ghost \cite{FZ,FGvN,LvN} and in addition, an ordinary massless vector of helicity $\pm1$ \cite{Rieg}. Analogous properties 
for conformal supergravity where recently explored in \cite{FKL1}.

The Lagrangian structure for higher $\N$-extended conformal supergravity was exhibited in \cite{BdRdW}, and lately
investigated in \cite{FKL2} based on the recent results  of \cite{Butter}, where the possible introduction of a holomorphic prepotential in $\N=4$ 
was discovered as suggested in \cite{BW}. Results on conformal supergravity \cite{FvP} related to the ghost structure,  its possible role in (supergravity) anomalies  \cite{Duff,FTs2,RvN}
and its higher $\N$ (extended conformal supergravity) extension include 
\cite{dWvH,LvN,LTs,KK,AG,FFKL,Salvio,Ts3,Ts4,FKvP,FL}.  Application of the superconformal formulation to cosmological models has been worked out in 
\cite{KL1,KL2,KL3}.

The  Einstein-Weyl theory \cite{Stelle} is described by the action
\begin{eqnarray}
S=\int_{\cal M} d^4 x\sqrt{-g}\Big{(}M_P^2 R+\frac{1}{2g_W^2} W_{\mu\nu\rho\sigma}W^{\mu\nu\rho\sigma}\Big{)}, \label{EW1}
\end{eqnarray}
where  $W_{\mu\nu\rho\sigma}$ is  the Weyl tensor defined as (in 4D)
\begin{eqnarray}
W_{\mu\nu\rho\sigma}=R_{\mu\nu\rho\sigma}+
g_{\mu[\sigma}R_{\rho]\nu}+g_{\nu[\rho}R_{\sigma]\mu}+\frac{R}{3} g_{\mu[\rho}g_{\sigma]\nu}.
\end{eqnarray}
Under conformal transformations of the metric 
\begin{eqnarray}
g_{\mu\nu}\to \widehat{g}_{\mu\nu}=\Omega^2 g_{\mu\nu}, \label{conf}
\end{eqnarray}
the Weyl tensor is invariant 
\begin{eqnarray}
\widehat{W}^\mu_{~\nu\rho\sigma}= W^\mu_{~\nu\rho\sigma} ,
\end{eqnarray}
and therefore, the second term in the action (\ref{EW1}) is invariant under conformal transformations 
\begin{eqnarray}
\sqrt{-\widehat g}\widehat{W}^{\mu\nu\rho\sigma}
\widehat{W}_{\mu\nu\rho\sigma}=
\sqrt{- g}W^{\mu\nu\rho\sigma}
W_{\mu\nu\rho\sigma},
\end{eqnarray}
in four dimensions. 
However, the full action (\ref{EW1}) fails to be  conformal invariant due to the Einstein-term  which transforms as 
\begin{eqnarray}
\widehat{R}= \Omega^{-2}R - 6 \Omega^{-3} g^{\mu\nu}\nabla_\mu 
\nabla_\nu \Omega,  
\end{eqnarray} 
so that it  breaks explicitly conformal invariance.
Hence, the Einstein-term plays the role of  mass term in this theory and can be regarded as  a mass deformation.

The propagator  of the Einstein-Weyl theory  as follows from the 
action (\ref{EW1}) 
 can be written as
\begin{equation}
\Delta_{\mu\nu\rho\sigma}=\Delta(k)  P_{\mu\nu\rho\sigma},
\label{propagator}
\end{equation}
where 
\begin{eqnarray}
\Delta(k)&=&{g_W^2\over k^2(k^2-g_W^2M_P^2)}, \label{d}\\
P_{\mu\nu\rho\sigma}&=&\frac{1}{2}\Big(\theta_{\mu\rho}\theta_{\nu\sigma}+\theta_{\mu\sigma}\theta_{\nu\rho}\Big)-\frac{1}{3}
\theta_{\mu\nu}\theta_{\rho\sigma}\, , ~~~ \label{P}
\end{eqnarray}
and 
\begin{eqnarray}
\theta_{\mu\nu}=\eta_{\mu\nu}-\frac{k_\mu k_\nu}{k^2}
\end{eqnarray}
 is the usual transverse  vector projection operator. Notice that   (\ref{propagator})  exhibits the  conformal $1/k^4$ behavior for $M_P^2=0$ (i.e., pure Weyl theory). For finite $M_P$, we  can  write $\Delta(k)$ equivalently  as
 \begin{eqnarray}
\Delta(k)=-\frac{1}{M_P^2}\frac{1}{k^2}+
 \frac{1}{M_P^2}\frac{1}{k^2-g_W^2 M_P^2},\label{pp0}
  \end{eqnarray}
  where the propagation of the  helicity-$\pm 2$ massless graviton is  easily recognized  in the first term of (\ref{pp0}). In addition,   there is also a massive spin-2 state associated to the second term of (\ref{pp0}). This massive state has a mass  given by the pole at $k^2=g_W^2 M_P^2$. However,  the residue at this pole  is  opposite  to that of the  massles graviton, and therefore it describes a ghost spin-2 state. This  shows that the theory contains as propagating degrees of freedom (dof) the standard,  massless spin-2 graviton $g_{\mu\nu}$ plus an additional massive spin-2 field $w_{\mu\nu}$.

In terms of helicity, the spectrum contains a spin-2 ghost $(s=2\to \pm 2,\pm1,0)$ and an ordinary massless ($\lambda=\pm 2$) massless graviton (a total of $7$ dof).  
This counting has a nice extension in the case of superconformal case. Actually to show that the Einstein-Weyl action (\ref{EW1}) describes both massless and massive spin-2, it is convenient to make a particular bimetric gravity with two spin-2 fields $g_{\mu\nu}$ and $w_{\mu\nu}$ with a new action 
\cite{Bergshoeff:2009hq}:
\begin{eqnarray}
S=\int_{\cal M} d^4 x\sqrt{-g}\Big{(}M_P^2 R(g)+{2M_P}G_{\mu\nu}(g)w^{\mu\nu}-M_W^2(w^{\mu\nu}w_{\mu\nu}-w^2)\Big{)}, \label{bimetric}
\end{eqnarray}
where 
\begin{eqnarray}
G_{\mu\nu}=R_{\mu\nu}-\frac{1}{2} R\,g_{\mu\nu}
\end{eqnarray}
 is the Einstein-tensor of the metric $g_{\mu\nu}$. The last term in (\ref{bimetric})  describes a mass term for the second spin-2 field
  $w_{\mu\nu}$. It is straightforward to see that (\ref{EW1}) and (\ref{bimetric}) are equivalent  \cite{Gording:2018not}.  Indeed, the equation of motion for $w_{\mu\nu}$ turns out to be
\begin{equation}
{\delta S\over \delta w_{\mu\nu}}\quad\Rightarrow\quad w_{\mu\nu}=\frac{M_P}{M_W^2}\left(R_{\mu\nu}(g)-{1\over 6}g_{\mu\nu}R\right)\, .\label{eqmotion}
\end{equation}
When the equation of motion is used in  (\ref{bimetric}), one gets the original action (\ref{EW1})  by using the identity 
\begin{eqnarray}
W_{\mu\nu\rho\sigma}W^{\mu\nu\rho\sigma}=GB+2(R_{\mu\nu}R^{\mu\nu}-\frac{1}{3}R^2), \label{GBW}
\end{eqnarray}
where 
\begin{eqnarray}
GB=R_{\mu\nu\rho\sigma}R^{\mu\nu\rho\sigma}-4R_{\mu\nu}R^{\mu\nu}+R^2
\end{eqnarray}
 is the Gauss-Bonnet tensor.   Therefore,  the actions (\ref{EW1}) and 
 (\ref{bimetric}) are equivalent up to total derivatives and the theory 
  contains one  massless and one massive spin-2  field $g_{\mu\nu}$ 
and $w_{\mu\nu}$, respectively. However, although the action for $g_{\mu\nu}$ is at non-linear level, the fields $w_{\mu\nu}$ appears just quadratically in (\ref{bimetric}).  Let us also note that although it looks that there is no kinetic term for the second field $w_{\mu\nu}$, the kinetic term of the latter is hidden  in  the term  $G_{\mu\nu}(g)w^{\mu\nu}$. Indeed, for small 
fluctuation $h_{\mu\nu}=g_{\mu\nu}-\eta_{\mu\nu}$ around Minkowski spacetime with flat metric $\eta_{\mu\nu}$, we find that, at quadratic level, the action (\ref{bimetric}) is written as 
\begin{eqnarray}
{\cal S}_2=\int d^4 x \sqrt{-g}\left\{-\frac{M_P^2}{2} h^{\mu\nu}{\cal E}^{\alpha\beta}_{\mu\nu} h_{\alpha\beta}+\frac{1}{g_W } w^{\mu\nu}{\cal E}^{\alpha\beta}_{\mu\nu} h_{\alpha\beta}- \Big(w^{\mu\nu}w_{\mu\nu}-w^2\Big)\right\}, \label{q2}
\end{eqnarray}
where  $g_W=M_W/M_P$ and ${\cal E}^{\alpha\beta}_{\mu\nu}$ is defined as
\begin{eqnarray}
{\cal E}^{\alpha\beta}_{\mu\nu} h_{\alpha\beta}=-\frac{1}{2}\left\{\Box h_{\mu\nu}-2\partial_{(\mu}\partial_\alpha h_{\nu)}^\alpha+\partial_\mu\partial_\nu
h-\eta_{\mu\nu}\Big(\Box h-\partial_\alpha\partial_\beta h^{\alpha\beta}\Big)\right\}.  
\end{eqnarray}
Then, in terms of the new field 
\begin{eqnarray}
\overline h_{\mu\nu}=h_{\mu\nu}-\frac{2}{g_W M_P^2} w_{\mu\nu}, 
\end{eqnarray}
the quadratic action (\ref{q2}) is written as 
\begin{eqnarray}
{\cal S}_2=\int d^4 x \sqrt{-g}\left\{-\frac{M_P^2}{2} \overline h^{\mu\nu}{\cal E}^{\alpha\beta}_{\mu\nu} \,\overline h_{\alpha\beta}+
\frac{1}{g_W^2 M_P^2 } w^{\mu\nu}{\cal E}^{\alpha\beta}_{\mu\nu} w_{\alpha\beta}- \Big(w^{\mu\nu}w_{\mu\nu}-w^2\Big)\right\}. 
\end{eqnarray}
Clearly now, the field $\overline h_{\mu\nu}$ is a massless graviton 
and $w_{\mu\nu}$ is a massive spin-2 tensor field. However, the kinetic term of  $w_{\mu\nu}$ has a wrong sign (compared to $\overline h_{\mu\nu}$) and therefore, it is a ghost  field.

\vskip0.2cm
\noindent
There are three interesting limits of the theory one may consider:

 \vskip0.2cm
\noindent{\sl (A) Decoupling of gravity:}

\noindent
First we consider  the infinite mass limit
\begin{equation}
 M_P\rightarrow \infty
 \, , \quad g_W={\rm fixed}
 \,.
\end{equation}
This is the limit where gravity decouples from the theory as this is the limit where gravity becomes
non-dynamical. In particular, if the coupling $g_W$ is finite,  
then both spin-2 states decouple completely, since in this case, the mass $M_W=g_W M_P$ of  $w_{\mu\nu}$ is infinite.

\vskip0.2cm
\noindent{\sl (B) Massless bigravity limit:}

\noindent
The limit of vanishing Planck mass  with finite $g_W$ 
\begin{equation}
 M_P\rightarrow 0
 \, , \quad g_W={\rm fixed}
 \, 
\end{equation}
is the massless limit of the theory.
In this limit, the propagator $\Delta(k)$ turns out to be 
\begin{equation}
\Delta(k)\rightarrow {g_W^2\over k^4}\, , \label{d1}
\end{equation}
in accordance with  conformal invariance.
Hence, the  spin-2 state $w_{\mu\nu}$  turns out to be  massless. This is a Higgs effect since for  finite $M_P$ 
the spin-2 state $w_{\mu\nu}$ becomes massive.  The theory in this limit is then the  massless Weyl gravity with action 
\begin{eqnarray}
S= {1\over 2 g_W^2}\int d^4 x\sqrt{-g} \, W_{\mu\nu\rho\sigma}W^{\mu\nu\rho\sigma}\, . \label{W2}
\end{eqnarray}
\noindent
This theory possesses conformal invariance and it propagates the following degrees of freedom:
 \vskip0.1cm
\noindent
(i) The standard  massless graviton $g_{\mu\nu}$, associated to  plane-waves 
in Einstein theory.

\vskip0.1cm
\noindent
(ii)   A massless spin-2 ghost state  $w_{\mu\nu}$ in the non-standard sector  corresponding to  non-planar waves.  There is also a massless  vector $w_\mu$, coming  from the 
$\pm 1$ helicities of the  massive $w_{\mu\nu}$ \cite{Rieg}. Notice that, due to conformal invariance, the helicity-0 state can be gauged away and it do not appear in the physical spectrum. 

 \vskip0.2cm
\noindent{\sl (C) Light spin-2 limit:}


\noindent
One may also consider  the double-scaling limit 
\begin{equation}
 M_P\rightarrow \infty\quad{\rm and}\quad g_W\rightarrow 0\, \quad{\rm with}\quad M_W<<M_P\, ,
\end{equation}
where $g_W$ vanishes  faster than $M_P^{-1}$. In this limit,  
 the standard gravitational 
sector (where the massless graviton $g_{\mu\nu}$ resides)  decouples from the  non-standard massless spin-2 ($w_{\mu\nu}$)  sector and the theory again describes massless Weyl with action (\ref{W2}). 
Indeed,  the propagator in Eq.(\ref{pp0}) can be approximated for 
$M_W=g_WM_P\to 0$  by 
\begin{eqnarray}
\Delta(k)&=&-\frac{1}{M_P^2}\frac{1}{k^2}+
 \frac{1}{M_P^2}\frac{1}{k^2-M_W^2}\nonumber \\
 &\approx& 
 -\frac{1}{M_P^2}\frac{1}{k^2}+\frac{1}{M_P^2}\frac{1}{k^2}
 \left(1+\frac{M_W^2}{k^2}+{\cal{O}}(M_W^4)\right)\approx 
 \frac{g_W^2}{k^4}+{\cal{O}}(M_W^4), \label{pp1}
\end{eqnarray}
which  is the propagator (\ref{d1}) for the massless Weyl gravity theory.

\section{The $\N$-extended 4D Superconformal Algebra}

For massless theories, it has been shown \cite{HLS} that the conformal group can be extended to the superconformal group which is generated by:
\begin{itemize}
\item The generators $(M_{\mu\nu},P_\mu,K_\mu,D$) of the conformal algebra 
$SO(2,4)=SU(2,2)$, i.e., the generators of the Poincar\'e algebra, the special conformal boosts $K_\mu$ and the dilation $D$ ,
\item the $2{\cal N}$ spinor generators $Q_\alpha^i, S_{\alpha i}, ~~(i=1,\cdots,{\cal N})$,
\item and ${\cal N}^2$ bosonic generators $R,T^i_j$, ($T^i_i=0)$
\end{itemize} 
with  (anti)commutation relations
\allowdisplaybreaks
\begin{eqnarray}
&&\{Q_\alpha^i,\overline Q_{\dot \alpha j}\}=2\delta^i_j \, \sigma^\mu_{\alpha \dot \alpha}\, P_\mu,~~~\{S_{\alpha j},\overline S_{\dot \alpha }^i\}=2\delta^i_j \, \sigma^\mu_{\alpha \dot \alpha}\, K_\mu,\nonumber \\
&&\{Q_\alpha^i,S_{\beta j}\}=- \delta^i_j \, \sigma^{\mu\nu}_{\alpha\beta}\, 
M_{\mu\nu}-4 \epsilon_{\alpha\beta} T^i_j-2 \epsilon_{\alpha\beta}\Big(R+iD\Big),\nonumber \\
&&[R,Q^i_\alpha]=\frac{4-N}{2N} Q^i_\alpha, ~~~[R,S_{\alpha i}]=-\frac{4-N}{2N} S_{\alpha i},\nonumber \\
&&[D,Q^i_\alpha]=\frac{i}{2} Q^i_\alpha, ~~~[R,S_{\alpha i}]=-\frac{i}{2} S_{\alpha i},\nonumber\\
&&[T^i_j, Q^k_\alpha]=-\frac{1}{2}\left(\delta^k_j Q^i_\alpha-\delta^i_j Q^k_\alpha\right), \nonumber \\
&&[T^i_j,S_{\alpha k}]=\frac{1}{2}\left(\delta^i_k S_{\alpha i}-\delta^i_j 
S_{\alpha k}\right).  \nonumber 
\end{eqnarray}
The bosonic generators $T^i_j$ and $R$ 
generate $SU({\cal N})$ transformations and  axial rotations on the spinors,
respectively. In particular,  there are no in general  central charges, except for  ${\cal N}=4$ where $R$ is central.
These relations, together with the standard commutation relations of the Poincar\'e subalgebra form the superconformal algebra $SU(2,2|{\cal N})$, which is the supersymmetric extension of the conformal algebra $SU(2,2)$. 
  
The standard superspace for the realization of the $SU(2,2|{\cal N})$ superconformal algebra is the coset 
\begin{eqnarray}
\mathbb{R}^{4|4{\cal N}}=\frac{SU(2,2|{\cal N})}{\{K,S,\overline S,M,D,T,R\}}=\big(x^\mu,
\theta_i^\alpha,\overline \theta^{\dot \alpha\, i}\big),
\end{eqnarray}
which  coincides with the  $\mathbb{R}^{4|4{\cal N}}$ superspace of  the Poincar\'e superalgebra. Superfields $\Phi(x^\mu,
\theta_i^\alpha,\overline \theta^{\dot \alpha\, i})$ are
superconformal primaries satisfying 
\begin{eqnarray}
S_{\alpha i}\Phi=\overline S ^i_{\dot \alpha}\Phi=K_\mu \Phi=0.
\end{eqnarray}
The UIR's of $SU(2,2|{\cal N})$ are then labeled as 
\begin{eqnarray}
D\Big(j_1,j_2,d,r,\{\vec{a}\}\Big),
\end{eqnarray}
where $d$ is the conformal weight, $(j_1,j_2)$ specify the Lorentz representation, $r$ is
 the $R$-charge and   $\{\vec{a}\}$ are the Dynkin labels of $SU({\cal N})$. 
Similarly to the super-Poicar\'e algebra, $SU(2,2|{\cal N})$ can be realised in a smaller superspace by using chiral supefields $\Phi(x,\theta)$ which  are elements of the coset 
\begin{eqnarray}
\mathbb{C}^{4|2{\cal N}}=\frac{SU(2,2|{\cal N})}{\{K,S,\overline S,M,D,T,
R,\overline Q\}}=\big(x^\mu,
\theta_i^\alpha\big). 
\end{eqnarray}
Such fields should satisfy 
\begin{eqnarray}
\overline Q_{\dot \alpha i}\Phi=0,
\end{eqnarray}
and the compatibility condition 
\begin{eqnarray}
\{\overline Q_{\dot \alpha i},\overline S ^j_{\dot \beta}\}\Phi=0,
\end{eqnarray}
gives that the chiral primary $\Phi$ is a Lorentz singlet ($(j_1=j_2=0)$ with 
$
d=-r. 
$

Let us note that the conformal algebras in $D=3,4,5$ and $D=6$ are
$Sp(4,\mathbb{R}),~ SU(2,2),~ SO(5,2)$ and $SO^*(8)=SO(6,2)$, respectively.  
These algebras are subalgebras of the bosonic subgroup of the superconformal groups in the corresponding dimensions. The latter are simple superalgebras
which are supersymmetry algebras and have been classified in \cite{nahm}. We have tabulated the simple superalgebras in the following table 1. 

\begin{center}
 \begin{tabular}{||c |c |c |c||} 
 \hline
 $D$ & Superalgebra & Bosonic subalgebra & Fermionic reps.\\ [0.5ex] 
 \hline\hline
 3 & $Osp(3|{\cal N})$ & $Sp(4,\mathbb{R})\times SO(N)$ & $(\underline{4},\underline{{\cal N}} $)\\ 
 \hline
 4& $SU(2,2|{\cal N})$ & $SU(2,2)\times SU({\cal N})\times U(1)$  & $(\underline{4},\underline{{\cal N}}) \oplus(\underline{\bar4},\underline{\bar {\cal N}}) $, ~~${\cal N}\neq 4$ \\
 \hline
 4& $SU(2,2|4)$ & $SU(2,2)\times SU(4)$  & $(\underline{4},\underline{4}) \oplus(\underline{\bar4},\underline{\bar 4}), $  ~~${\cal N}=4$ \\
 \hline
 5 & $F(4)$ & $SO(5,2)\times SU(2)$ & $(\underline{8},\underline{2})$ \\
 \hline
 6 & $Osp(6|{\cal N})$ & $SO(8^*)\times USp(2{\cal N})$& $(\underline{8},\underline{2{\cal N}})$ \\
 \hline
\end{tabular}
\end{center}
We observe from the above table that the bosonic subalgebra in $D$-dimensions contains the conformal group in that dimension and the fermionic  generators
are in the spinorial representation of the conformal group. In fact, this is the condition for a superalgebra to be a supersymmetry algebra, namely, the odd generators should transform in the spinor represenation of the even subalgebra of the superalgebra.
 
The conformal supermultiplets can be realized in Lagrangian theories on  chiral and vector multiplets for ${\cal N}=1$,  hyper and vector multiplets 
for ${\cal N}=2$ and vector multiplets for ${\cal N}=3,4$. However, when 
coupled to Weyl supergravity, ${\cal N}=3,4$ are different because massive spin-2 multiplets are different in ${\cal N}=3$ and ${\cal N}=4$ (comprised of $128$ and $256$ states, respectively). 
Note that superconformal Lagrangians can be realized only for 
${\cal N}\leq 4$. This  bound is in analogy with the ${\cal N}\leq 8$ bound of 
Poincar\'e supergravity in $D=4$. The corresponding bound $D\leq 11$ for Poincar\'e supergravity corresponds to $D\leq 6$ for the higher dimensional conformal supergravity. The bound ${\cal N}\leq 4$ can be understood as the bound of spin-2 massive supermultiplets in extended supesymmetry 
\cite{deWit:1978pd}.

Another way to understand the ${\cal N}\leq 4$ in $4D$ is to compute the quadratic part of the super-Weyl action using the superconformal Ward identities \cite{FK1}. It turns out that the Weyl term and the $U({\cal N})$ 
gauge fields occur as follows
\begin{eqnarray}
  {\cal L}=2\alpha \left(1-\frac{4}{{\cal N}}\right) F^{\mu\nu}F_{\mu\nu}
  -4\alpha F^{i\mu\nu}F^i_{\mu\nu}+8\alpha W^{\mu\nu\rho\sigma}W_{\mu\nu\rho\sigma},
  \end{eqnarray}  
where $F_{\mu\nu}$ and $F^i_{\mu\nu}$ ($i=1,\cdots,{\cal N}$)  are the field strengths of the $U(1)$ and $SU({\cal N})$ subgroups of $U({\cal N})$.   Clearly, only for ${\cal N}\leq 4$ the $U({\cal N})$ gauge fields have the same sign and so they can belong to the same multiplet. For ${\cal N}=4$, the $U(1)$ gauge field has no kinetic term.  
 
\section{Massless Higher Derivative Theories} 

Let us apply canonical quantization to the (dipole ghost) Lagrangian
\begin{eqnarray}
 {\cal L}_A=\frac{1}{2}\Big(\partial_\mu\partial_\nu A)^2,
 \end{eqnarray} 
 which is equivalent to 
 \begin{eqnarray}
  {\cal L}_A= A_1\Box A-\frac{1}{2}\lambda^2 A_1^2,  \label{LB}
  \end{eqnarray} 
  with $A_1$ a Lagrange multiplier field. The equations of motion for the fields 
  $A$ and $A_1$  are 
  \begin{eqnarray}
   \Box A=\lambda^2 A_1, ~~~\Box A_1=0.
   \end{eqnarray} 
We can construct the Fock space as usual: we introduce creation and annihilation operators $a(k), a_1(k)$ and $a^\dagger(k),a_1^\dagger(k)$ respectively, where the vacuum is defined as 
\begin{eqnarray}
a(k)|0\rangle=a_1(k)|0\rangle=0.
\end{eqnarray}
The one-particle states are 
 \begin{eqnarray}
 |k\rangle=a^\dagger(k)|0\rangle, ~~~|k\rangle_1=a_1^\dagger(k)|0\rangle, 
 \label{g1}
 \end{eqnarray}
and they satisfy 
\begin{eqnarray}
 {}_1\langle k|k'\rangle_1=\langle k|k'\rangle=0, ~~~~{}_1\langle k|k'\rangle=\delta^{(3)}_D(k-k').   \label{g2}
 \end{eqnarray} 
 Only one of these one-particle states is an energy eigenstate
 \begin{eqnarray}
 P^0|k\rangle_1=\omega |k\rangle_1, ~~~~P^0|k\rangle=\omega |k\rangle+\frac{\lambda^2}{2\omega}|k\rangle_1.
 \end{eqnarray}
  Eqs.(\ref{g1}) and (\ref{g2}) are typical of a dipole ghost state.
  By using a linear transformation, the two states $|k\rangle$ and $|k\rangle_1$ can be diagonalized such that, one of them will have positive and one negative norm but none of them would be then an energy eigenstate.

For the fermions we have the Lagrangian 
\begin{eqnarray}
 {\cal L}_\psi=\partial_\mu \overline \psi \slashed \partial \partial^\mu \psi,
 \end{eqnarray} 
 which can be written, with the help of  Lagrange-multipliers fermion fields
 $\psi_1,\psi_2$ as 
 \begin{eqnarray}
  {\cal L}_\psi= \overline \psi_1 \slashed \partial  \psi_1+
  \overline \psi \slashed \partial  \psi_2+\lambda \overline \psi_2 \psi_1. 
  \end{eqnarray} 
  Note that adding a second scalar $B$ with the same Lagrangian (\ref{LB}) 
  (or similarly $B$ and $B_1$) and  two extra scalars $F$ and $G$ with standard kinetic terms, the full theory with the fields $(A, A_1, B, B_1, \psi, 
  \psi_1, \psi_2, F, G)$ is supersymmetric \cite{FZ}. In particular, the complex scalar 
  $(A+iB)$ is a dipole ghost and an ordinary scalar, whereas, the Majorana 
  (or Weyl) fermions form a tripole ghost fermion, that is a dipole ghost and
   an ordinary  fermion. Only two of the one-particle states of a triple ghost $(|k\rangle,
   |k\rangle_1,|k\rangle_2)$ are energy eigenstates and they satisfy
   \begin{eqnarray}
    P^0|k\rangle_1=\omega |k\rangle, ~~~P^0|k\rangle_2=\omega |k\rangle_2, ~~~P^0|k\rangle=\omega |k\rangle+\frac{\lambda^2}{2\omega} |k\rangle_2. 
    \end{eqnarray} 
    The theory describes $6_B+6_F$ dof. The $6_B$ bosonic dof correspond to 1 dipole ghost and 1 complex scalar, whereas the $6_F$ fermionic dof are associated with the dof of the triple ghost, that is a dipole ghost ($\psi,\psi_2)$ and an ordinary fermion ($\psi_1$). This is the spectrum of  a supersymmetric theory 
describing one dipole WZ ghost multiplet and one ordinary WZ physical multiplet.

\section{From Rigid to Local Supersymmetry}

As we have discussed above, the massive Weyl gravity, given in Eq.(\ref{EW1}) propagates $7\!=\!2\!+\!5$ dof associated to a massless helicity $\pm2$ graviton states, and $1$ massive spin-2 state with mass   $g_w M_p$ 
and $5\!=\!(\pm 2,\pm 1,0)$ dof. In the massless limit, the Weyl symmetry removes the scalar dof and there remain $6\!=\!(\pm 2+\pm 2+\pm1)$ states where the $\pm2$ helicities made up a dipole ghost, and the $\pm1$ helicity corresponds to an ordinary vector. 

\subsection{${\cal N}=1$ Massive and Massless Weyl Supergravity}
The central superfield in 
 conformal supergravity is the super-Weyl tensor ${\cal W}_{\alpha\beta\gamma}$, where $(\alpha,\beta,\gamma=1,2)$ are  $SL(2,\mathbf{C})$ spinor indices. It is  chiral superfield with  spin $({3\over 2},0)$, so that  its highest component has  spin-3/2. In addition,  
 ${\cal W}_{\alpha\beta\gamma}$ accommodates also the five dof of a massive spin-2 field.  
The Weyl superfield ${\cal W}_{\alpha \beta \gamma}$ is chiral so that it satisfies the conditions
\be
{\cal W}_{\alpha \beta \gamma} = {\cal W}_{(\alpha \beta \gamma)} \  , \  
({\cal W}_{\alpha \beta \gamma})^* = \overline {\cal W}_{\dot \alpha \dot \beta \dot \gamma} \  , \   
\overline{ {\cal D}}_{\dot{\delta}} 
{\cal W}_{\alpha \beta \gamma}= 0  \ ,  \label{WW}
\ee
and it has in its components  
the Weyl tensor $W_{\mu\nu\rho\sigma}$  and the vector auxiliary $A_\mu$ 
of the ${\cal N}=1$ supergravity \cite{FFKL}.

 \subsubsection{Massive theory}

%
The bosonic part of the massive supesymmetric   Weyl theory turns out to be 
\cite{CFS,FFKL} 
\be
\nonumber
e^{-1} {\cal L}_{{\cal W}} =  \frac{1}{2}R+\frac{1}{3} A_\mu A^\mu-\frac{1}{3} u\overline u+
\frac{\alpha}{2} 
\left( \frac12 R^2_{HP}
-\frac{2}{3} F_{\mu\nu} F_{\rho\sigma} \epsilon^{\mu\nu\rho\sigma}
\right) +
\frac{1}{2g^2}  \left( W^{\mu\nu\rho\sigma}W_{\mu\nu\rho\sigma}
- \frac43 F^{\mu\nu} F_{\mu\nu}
\right) 
\, , 
\ee
where  $u$ is the auxiliary scalar,  $F_{\mu\nu}=\partial_\mu A_\nu-\partial_\nu A_\mu$ is the field strength of $A_\mu$ and  $R^2_{HP} = W_{\mu\nu \kappa\lambda} \epsilon^{\kappa\lambda \rho\sigma} W_{\rho\sigma}^{\ \ \mu\nu}$ is the Hirzebruch--Pontryagin  tensor.  It is clear that the auxiliary vector of ${\cal N}=1$ Poincar\'e supergravity acquires a kinetic term and therefore it is   dynamical in conformal supergravity. 
The  spectrum of the massive ${\cal N}=1$ super-Weyl theory  
contains $n_B+n_F=20$ dof in the following form:
\vskip0.1cm
\noindent
(i) The  standard massless spin-2 graviton multiplet $(h,q_R)$ with $n_B+n_F=4$ dof,  helicity $h$ and 
$q_R$ $U_R(1)$ charge: 
\begin{eqnarray}
g_{{\cal N}=1}:~   (+2,0) +  (+{3\over2},+{1\over 2} ) +{\rm CPT}   \, .
\end{eqnarray}
\vskip0.1cm
\noindent
(ii)  The massive spin-2 supermultiplet in the non-standard sector with 
 $n_B+n_F=16$ dof which form the  
massive ${\cal N}=1$ super-Weyl multiplet  \cite{FZ}:
\begin{eqnarray}
w_{{\cal N}=1}:~    
 \Big(2,2\Big(\frac{3}{2}\Big),1\Big) \, .
\end{eqnarray}
It should be noted that  the massive states in ${\cal N}=1$ 
supergravity form  $USp(2)$ representations.
\vskip0.2cm
\noindent

\subsubsection{Massless theory}

The spectrum of the massless ${\cal N}=1$ super-Weyl theory is the following:
\vskip0.1cm
\noindent
(i) A  standard massless spin-2 $(h,q_R)$ supergravity multiplet with $n_B+n_F=4$ dof
\begin{eqnarray}
g_{{\cal N}=1}:~    (+2,0) +   (+{3\over2},+{1\over 2}) +{\rm CPT}   \, .
\end{eqnarray}
\vskip0.1cm
\noindent
(ii)  In the non-standard sector,   
 we have     a massless ghost-like spin-2 supermultiplet, arising  from the massive Weyl multiplet  $w_{{\cal N}=1}$: 
\begin{eqnarray}
{\rm spin-2}:~    (+2,0) +   ( +{3\over2},+{1\over 2} )+{\rm CPT},~~~(\mbox{4 dof})  \, .
\end{eqnarray}
In addition, the $w_{{\cal N}=1}$ multiplet gives a massless, physical spin-3/2 supermultiplet:
\begin{eqnarray}
{\rm spin-3/2}:~    (+{3\over2},-{1\over 2}) +  (+1,0) +{\rm CPT},~~~(\mbox{4 dof})   \, , \label{32v}
\end{eqnarray}
a massless spin-one vector multiplet:
\begin{eqnarray}
{\rm spin-1}:~    (+1,0) +  ( +{1\over2},+{1\over 2})+{\rm CPT},~~~(\mbox{4 dof})         \label{v11}
\end{eqnarray}
and a chiral spin-1/2 multiplet:
  \begin{equation} 
  {\rm spin-1/2}:~ (+{1\over 2},-{1\over 2}) +  ( 0,0)+{\rm CPT},~~~(\mbox{4 dof})       \, . 
  \end{equation}
Note that the spin-3/2 fields form 
a 
tripole ghost. The latter acts effectively  as a physical spin-3/2 multiplet together with  a  spin-2 dipole ghost multiplet \cite{FZ}.
However, the chiral  multiplet is unphysical as it can be gauged away due to the superconformal and $U(1)_R$  symmetries \cite{FKL1}.

\vskip0.2cm
\noindent
Recapulating, the massless super-Weyl gravity theory contains $n_B+n_F=16$ physical, propagating degrees of freedom \cite{LvN}.
In addition, the gravitino action contains 
(${\cal N}=1$) eight helicity states $(+\frac{3}{2},+\frac{1}{2}),(+\frac{3}{2},+\frac{1}{2}),(+\frac{3}{2},-\frac{1}{2})$ 
(+  CPT conjugates) \cite{LvN},  
which is important in anomaly cancellation. 
The counting of dof is summarized in the following table 2. 
\begin{center}
 \begin{tabular}{||c |c |c |c||} 
 \hline
 ${\cal N}$ & DoF of Massive Weyl & Gauged DoF in Massless Weyl  & DoF of Massless Weyl\\ [0.5ex] 
 \hline\hline
 0 & $5+2=7$ & 1 scalar & $7-1=6$ \\ 
 \hline
 1 & $16+4=20$ & 1 chiral multiplet & $20-4=16$    \\
 \hline
 2 & $48+8=56$ & 1 vector and 1 hyper & $56-16=40$  \\
 \hline
 3 & $128+16=144$ & 3 vector multiplets & $144-48=96$ \\
 \hline
 4 & $256+32=288$ & 6 vector multiplets & $288-96=192$ \\
 \hline
\end{tabular}
\end{center}
 
 In fact, a formula that gives the number of degrees of freedom in ``massless"  ${\cal N}$-extended supergravity is 
 
 \begin{eqnarray}
 &&{\cal N}=0,1,2,3, ~~~ {\rm dof}=(3+{\cal N}) \cdot 2^{{\cal N}+1};~~~~
 \mbox{ with matter (n)} ~~~{\rm dof}=(3+{\cal N}+n)\cdot 2^{{\cal N}+1},
 \nonumber \\
 &&{\cal N}=4,~~~~~~~~~~~~{\rm dof}=12 \cdot 2^4; ~~~~~~~~~~~~~~~
 \mbox{ with matter (n)} ~~~{\rm dof}=(12+n)\cdot  2^4,
 \end{eqnarray}
 without and with $(n)$ matter multiplets.
The factor $3+{\cal N}$ denotes the two (dipole) graviton multiplets, the  vector multiplet and the ${\cal N}$-gravitini multiplets (all massless). The
${\cal N}$-gravitini multiplets contain the ${\cal N}^2$ gauge bosons, which gauge the $U(N)$ symmetry. For ${\cal N}=4$, the extra vector multiplet is the partner (Weyl vector) of the Weyl graviton that for ${\cal N}<4$ seats in the extra vector multiplet. 

 \section{${\cal N}=4$ Weyl Supergravity and its Massless (Ghost) Spectrum}

As we have seen above, the dof of the massless ${\cal N}=4$  Weyl supergravity has $192$ dof which is the original $288$ of the massive theory (a massive and a massless spin-2) after gauging away $96$ gauge degrees of freedom of six vector multiplets due to superconformal invariance.   Then the remaining massless spectrum is given by $2$ ${\cal N}{=}4$ graviton multiplets ($2\times 32$ dof) and $4$ (spin-3/2) gravitini multiplets ($4\times 32$ dof) giving a total of $6\times 32=192$ dof. This allows to classify all the states and 
compare with the corresponding  Lagrangian \cite{BdRdW}. This reproduces the known structure which is cubic in certain fermions (spin-3/2) and 
(spin-1/2) and standard in others ($\overline{20}$) \cite{JJ}. 
 In particular, we have:
 \vspace{-3mm}
 \begin{itemize}
 \item {\it Bosonic Spectrum:} Form the spin-2 (dipole) graviton multiplets we have 
 \begin{eqnarray}
 2\times \Big[ (+2,\underline 1)+(+1,\underline 6)+(0,\underline1)+{\rm CPT}   \Big],
 \end{eqnarray}
 whereas from the $\overline{\underline4}$ gravitino multiplets we get 
 \begin{eqnarray}
 (+1,\underline{15}+\underline1)+(0,\overline{\underline{10}}+\underline6)+{\rm CPT}. 
 \end{eqnarray}
 The vectors and scalars in the $6$ correspond to the higher derivative graviphoton contribution, whereas the vectors in the $(\underline{15}+\underline1)$ correspond to the $SU(4)$ gauge fields and the Weyl vector. In particular, the scalars in the $\overline{\underline{10}}$ have standard two derivatives kinetic terms while the $SU(4)$ singlet has higher derivative (quartic) kinetic term. 
 \item {\rm Fermionic Spectrum:} From the two  (dipole) ${\cal N}=4$ gravitino multiplets we have 
 \begin{eqnarray}
 \Big(\!+\frac{3}{2},4\Big)+\Big(\!+\frac{3}{2},4\Big)+\Big(\!+\frac{1}{2},\overline{4}\Big)+\Big(\!+\frac{1}{2},\overline{4}\Big)+{\rm CPT},
 \end{eqnarray}
 \noindent
 and from the $\overline{\underline{4}}$ gravitino multiplets we get 
 \begin{eqnarray}
 \Big(\!+\frac{3}{2},\overline{\underline{4}}\Big)+\Big(\!+\frac{1}{2},\overline{\underline{20}}\Big)
 +\Big(\!+\frac{1}{2},\underline{4}\Big)+\Big(\!+\frac{1}{2},
 \underline{4}\Big)+{\rm CPT}.
 \end{eqnarray}
 \end{itemize}
 In addition, the gravitino and the spin-1/2 fields have cubic kenetic terms 
 whereas the spin-1/2 fields in the $\overline{\underline{20}}$ have physical standard kinetic terms.

\vskip.2in
\noindent
{\bf {Acknowledgment}}

\vskip.05in
\noindent
We would  like to thank A. Sagnotti for enlighting discussions.  
The work of S.F. is supported in part
by CERN TH Department and INFN-CSN4-GSS.  The work of D.L. is supported by the Origins Excellence Cluster.


\end{document}